\pdfoutput=1
\documentclass[aps,twocolumn]{revtex4}

\usepackage{dcolumn}
\usepackage{color}
\usepackage{graphicx}
\usepackage{amssymb}

\begin{document}

\title{Power-Law Scaling in Protein Synthesis of a Stochastic Regulon}

\author{Emily Chapman-McQuiston$^{1}$, Chuck Yeung$^{2}$, and X.L. Wu$^{1}$}

\date{July 28, 2008}

\affiliation{
$^{1}$Department of Physics and Astronomy, University of Pittsburgh, PA 15260 \\
$^{2}$School of Science, The Pennsylvania State University at Erie,
The Behrend College, Erie, PA 16583}

\begin{abstract}
We investigate the protein expression pattern of the \emph{lamB} gene
in \emph{Escherichia coli} LE392. The gene product LamB is an important
membrane protein for maltose transport into cells but it is also exploited
by bacteriophage $\lambda$ for infection. Although our bacterial
population is clonal, stochastic gene expression leads to a majority
population with a large receptor number and a minority population
with a small receptor number. We find that LamB receptor distribution
$p(n)$ of the small-$n$ population is scale invariant, $p(n)\sim n^{-\alpha}$,
where the exponent $\alpha$ depends on growth conditions. A heuristic
model is proposed that relates the observed exponent to the protein
production rate.
\end{abstract}

\maketitle
The occurrence of scale invariance in physics and biology often reflects
an important underlying principle\cite{Pascual,Volkov,McKane}. Herein,
we present a novel observation of spontaneous/stochastic gene expression
that gives rise to a power-law distribution $p(n)\propto n^{-\alpha}$,
where $n$ is the number of LamB receptors in individual bacteria.
The experiment is made possible by a dual-colored phage labeling technique
that renders the minority population quantifiable in a flow cytometer.
The LamB is a maltose channel, but it is also exploited by $\lambda$
phage as an infection site. We find that in addition to the main population
with an average $n\simeq500$, there is a small sub-population ($\sim1\%$)
with a small receptor number. This small population increases the
likelihood that the bacterial population as a whole survives a phage
attack\cite{Chapman-McQuiston-Thesis}. Surprisingly, the small-$n$
population has a scale-invariant distribution $n^{-\alpha}$, behaving
very differently from the log-normal distribution commonly seen for
major proteins in a bacterium\cite{Furusawa,Krishna}. We propose
a model that takes into account the rate of protein synthesis and
protein dilution due to cell division. This model yields a steady-state
distribution $p_{0}(n)\sim n^{-\alpha}$ with $\alpha$ being a continuous
function of the rate of protein synthesis.

Our experiments were carried out using \emph{E.\ coli} LE392 that are sensitive
to phage $\lambda$\cite{Chapman-McQuiston}. The maltose regulon of this strain
is inducible; the LamB receptor number $n$ can be varied from a few to
$\sim10^{3}$ depending on culture conditions\cite{Schwartz}. The bacteria were
grown in M9 minimal medium supplemented with either 0.4\% glucose or 0.4\%
maltose. Standard protocols were used to grow the bacteria and to purify the
$\lambda$ phage\cite{Chapman-McQuiston}. The \emph{lamB} is an essential gene
for maltose metabolism when the substrate is present at a very low level
($\ll10\,\mu M$). However, in the current experiment, the maltose level is
sufficiently high that the low receptor numbers does not impede the growth of
the minority population~\citep{Chapman-McQuiston,Chapon}. To quantify the
number of LamB receptors $n$ for individual cells, bacteria were incubated with
high concentrations of fluorescently labeled $\lambda$ phage (dye/phage ratio
$\gtrsim2000$), and the brightness of each bacterium was interrogated using
flow cytometry (Dako Cyan ADP). Using our labeling procedure, individual phage
particles were bright enough to reach the detection threshold of the flow
cytometer\cite{Chapman-McQuiston}, making them quantifiable by the instrument.

\begin{figure}

\includegraphics[width=8cm]{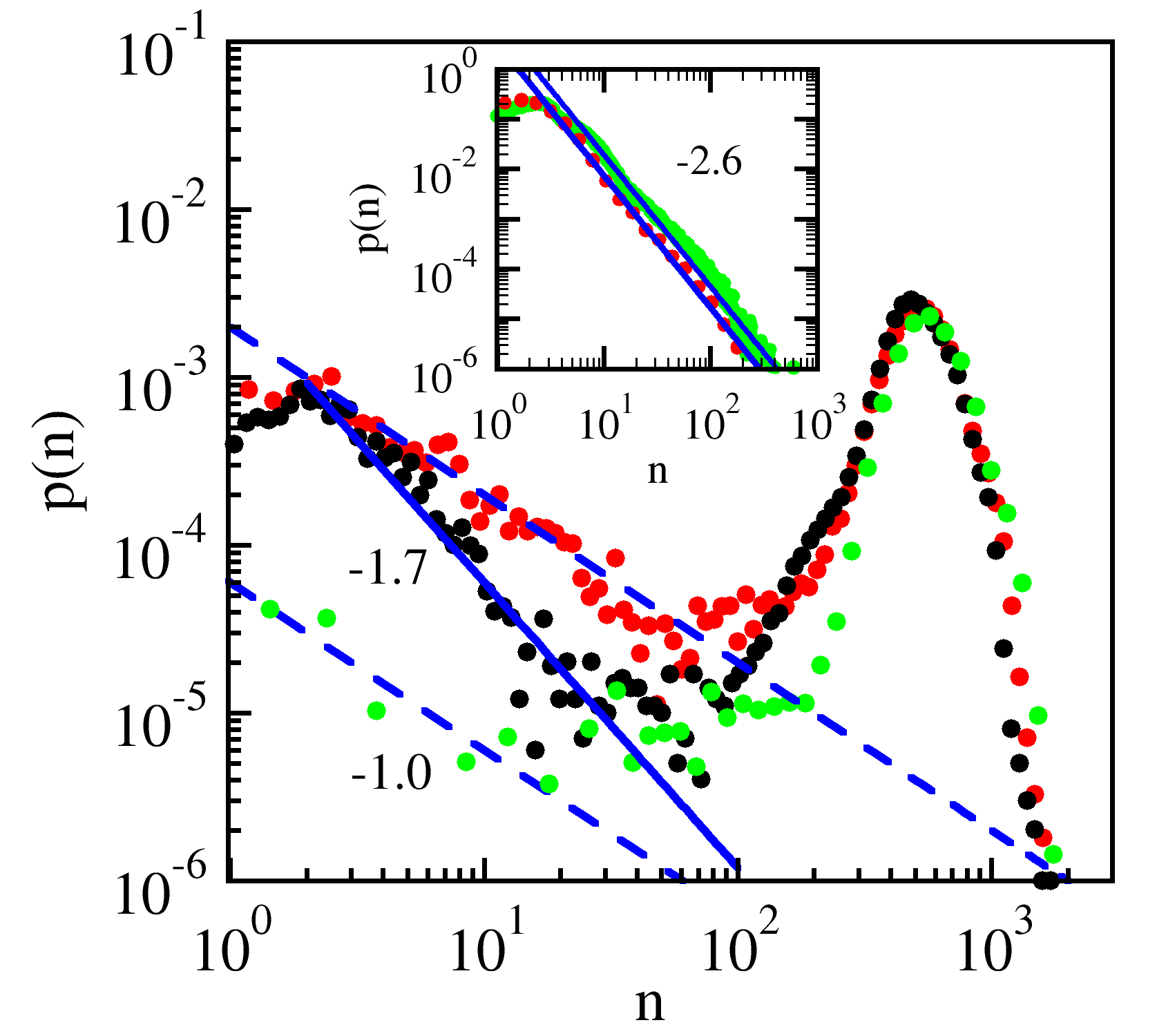}

\caption{Three steady-state distributions of $p_{0}(n)$ of bacteria grown
in M9+0.4\% maltose. The PDFs of the minority populations follow power
laws $p_{0}(n)\propto n^{-\alpha}$ with $\alpha=1$ for the dashed
lines and $\alpha=1.7$ for the solid line. The inset displays steady-state
distributions of bacteria grown in M9+0.4\% glucose. Since the maltose
regulon is repressed, there is no majority population. The exponent
for both runs was found to be $\alpha\simeq2.6$ as shown.
\label{fig:1}}

\end{figure}

We show several runs of LE392 grown in M9+maltose in Fig.\
\ref{fig:1}. The steady-state LamB probability distribution function
(PDF) $p_{0}(n)$ has two peaks, one around $n\sim500$ and the other
near $n\simeq0$. The minority peak appears prominently in the log-log
plot, but this sub-population is only a few percent of the total population.
In rare cases, the fraction is even smaller, as indicated by the lowest
curve in the figure. For all runs, the minority population has a power-law
distribution $p_{0}(n)\propto n^{-\alpha}$ that extends more than
a decade in $n$. The measured exponent $\alpha$ is not constant
but varies from run to run with $\alpha$ ranging from $1$ to $2$.
This variation may reflect the sensitivity of $\alpha$ to the varying
protein production rates in different bacterial cultures. On the other
hand, the PDF of the majority population is quite reproducible, and
has the usual log-normal form\cite{Krishna}.\textcolor{red}{ }\textcolor{black}{~We
conjecture that segregation of cells into minority and majority populations
is shaped by phage selection. The low $n$ phenotype allows the bacterial
population to hedge its bets due to the dual functionality of LamB
receptors. These minority cells are less fit when maltose is not abundant
in the environment, but their presence increases the chance that the
population as a whole survives a phage attack.}

We used a dual-colored phage labeling technique
to investigate the role of the minority cells in the recovery of a
bacterial population after a phage attack. Specifically, we are interested
in the time scale of recovery, and the relationship between the minority
and the majority populations. Our method consists of introducing
red-labeled (AlexaFluor 633) $\lambda$ phage into the bacterial population.
The bacteria is exposed to the phage for 30 minutes, and the phage
pressure, defined as the initial phage concentration $P(0)$, is then
relieved by several washing steps\cite{Chapman-McQuiston-Thesis}.
The phage selectively decimates bacteria that express a large number
of receptors, leaving only cells with small number of receptors. The
surviving bacteria were then allowed to regrow in a fresh medium.
Throughout the regrowth period, the bacteria were continuously diluted
into prewarmed fresh media. This procedure ensures that the bacterial
concentration is always less than $10^{7}\, cc^{-1}$, and the background
phage concentration is below $10^{-5}\, cc^{-1}$ so that the probability
that a phage binds to a bacterium is very small\cite{Chapman-McQuiston-Thesis}.
To quantify the time-dependent receptor distribution $p(n,t)$, a
small sample was periodically taken from the exponentially growing
culture. The receptors on individual cells were tagged with green-labeled
(AlexaFluor 488) $\lambda$ phage and interrogated via flow-cytometry\cite{Chapman-McQuiston}.
To stringently discriminate deceased cells, the bacterial sample was
also treated with propidium iodide, which specifically labels the
dsDNA of infected cells making them fluoresce brightly in orange.
The green channels of the flow cytometer is conditioned on the red
(the killing phage) and the orange channels (the dead cells) to get
rid of the background count.

\begin{figure}

\includegraphics[width=8cm]{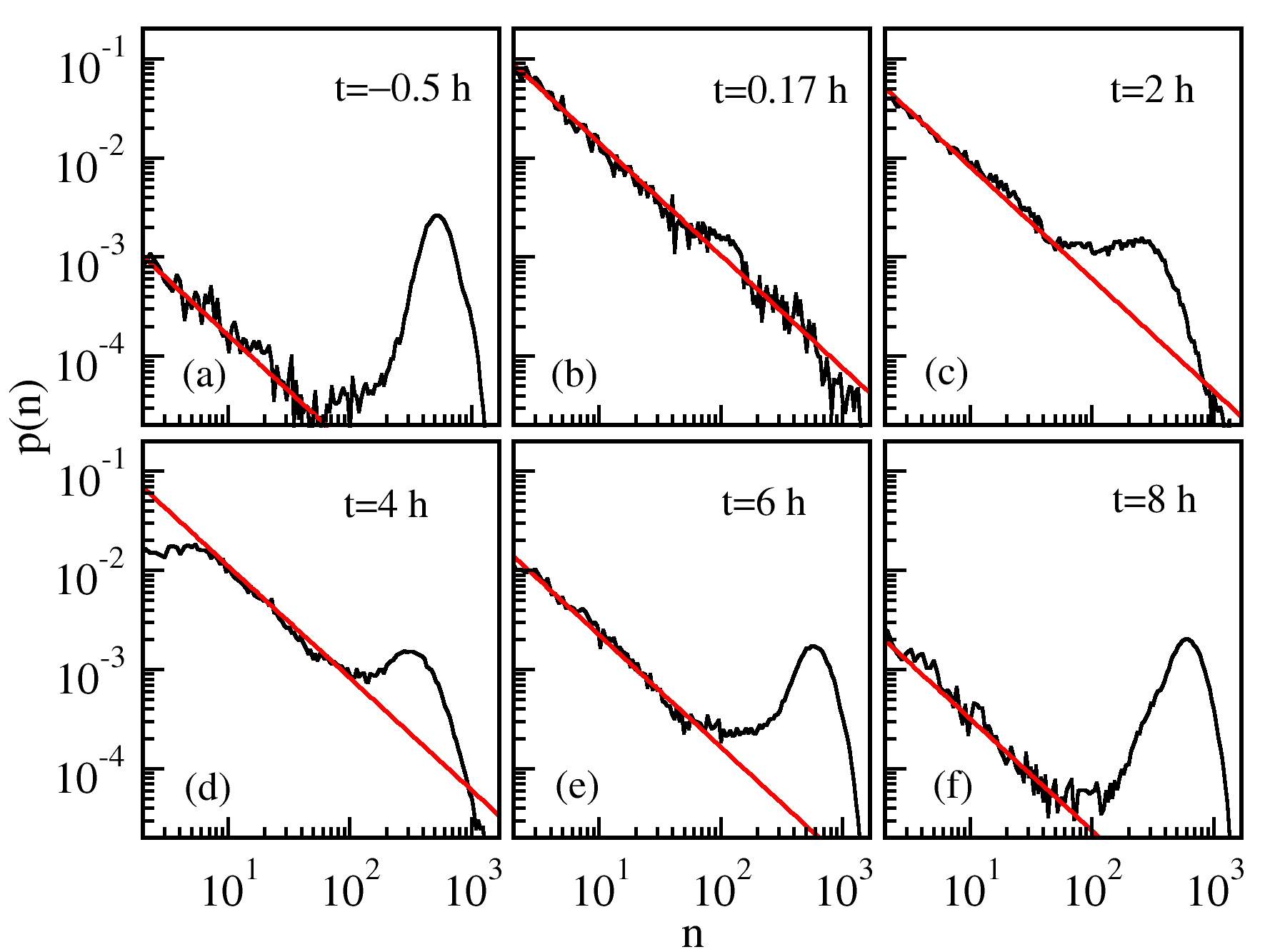}

\caption{
LE392 bacteria at an initial concentration of $1\times10^{7}~cm^{-3}$
and a $\lambda$ phage concentration $1\times10^{9}~cm^{-3}$ were
incubated at $T=37\,^{o}C$ for 30 minutes. The phage pressure was
then released by washing the cells twice using pre-warmed M9+0.4\%
maltose. This defines $t=0$ in the above plots. The population has
two peaks before phage is introduced (a). Immediately after phage
selection (b), the majority population is decimated, leaving only
the low $n$ distribution that still scales as $n^{-1}$ over three
decades in $n$. After the phage removal, a broad peak emerges from
the $n^{-1}$ background in $\sim2$hrs (c), and the peak sharpens
and moves to large $n$ with time (c-f). Even after 8 hrs (f), corresponding
to $\sim11$ generations, the majority population is still not fully
recovered as indicated by the lower height of $p(n)$.
\label{fig:2}
}

\end{figure}

Figure \ref{fig:2} depicts a sequence of LamB PDFs, $p(n,t)$, before and after
the bacteria were exposed to $\lambda$ phage. Before the phage exposure (Fig.\
\ref{fig:2}a), the LamB distribution has a primary peak at $n\simeq500$ and a
secondary peak at $n\sim0$ similar to the data in Fig.\ \ref{fig:1}. The
minority peak at small $n$ in this case scales as $p_{0}(n)\sim n^{-1}$ as
delineated by the solid lines in Fig.\ \ref{fig:2}a. The majority population is
quickly eliminated by the exposure to the $\lambda$ phage (Fig.\ \ref{fig:2}b),
but cells with a small $n$ remain unaffected. Here we observed that the $n^{-1}$
scaling extends over more than three decades $10^{0}<n<10^{3}$. After the
removal of phage, the majority cell population slowly grows back to its original
form in $\sim8$ hrs (Fig.\ \ref{fig:2}(c-f)), with the distribution of the
minority cells maintaining the same shape the entire time. This result is robust
in that the functional form of the minority population is time invariant for a
wide range of initial phage concentrations $P(0)$ so long as $P(0)$ is not large
enough to erode the minority population. This observation suggests that
stochastic switching between the minority and the majority states is relatively
slow, indicating a weak coupling between the two populations. In other words,
the minority population is able to replenish itself, maintaining the
self-similar form independent of the majority population.

The maltose regulon is inactivated if LE392 is grown in M9+glucose
instead of maltose. This is a part of catabolic repression resulting
from a sharp drop of cAMP concentration in cells when glucose is present.
However, stochastic events still lead LamB to be expressed at a very
low level $\bar{n}\simeq2$. The inset of Fig.\ \ref{fig:1} displays
measurements of the steady-state LamB distribution $p_{0}(n)$ for
this case. We observe that, over a broad range of $n$, $p_{0}(n)$
also decays as a power law with the exponent $\alpha\simeq2.7\pm0.2$.
In this case, the run-to-run variation of $\Delta\alpha\simeq0.2$
is smaller than that of the maltose-grown cells. Therefore, the uninduced
(or chemically repressed) bacteria and the minority population of
the induced bacteria both have power-law LamB distributions $p(n)\sim n^{-\alpha}$
although with somewhat different $\alpha$ and $\Delta\alpha$. The
larger $\Delta\alpha$ seen in the maltose-grown cells suggests that
this minority population is more sensitive to external variations
than uninduced cells and are hence more susceptible to environmental
changes.

The population-balance approach\cite{Ramkrishna} can be used to
help understand the protein distributions in a bacterial population.
The number of cells with $n$ proteins at time $t$, $B(n,t)$, obeys
a master equation:
\begin{widetext}
\begin{equation}
	\frac{\partial B(n,t)}{\partial t}
		=
	-\gamma(n)B(n,t)-\frac{\partial}{\partial n}(\nu(n)B(n,t))
		 +
	 2\intop_{n}^{\infty}\gamma(m)\Phi(n|m)B(m,t)~dm,
	 	\label{eq1:a}
\end{equation}
\end{widetext}
where $\Phi(n|m)$ is the partition probability of a daughter cell
inheriting $n$ receptors from a mother cell with $m$ receptors,
and $\nu(n)$ and $\gamma(n)$ are respectively the state-dependent
protein production rate and the cell division rate. The integration
on the right-hand side of Eq. \ref{eq1:a} takes into account the
long-range effect of protein redistribution as a result of cell division.
Based on our experiments, the division rate for the majority and minority
cells are about the same, and as an approximation, we assume $\gamma(n)=\gamma$.
In the steady state, a separation of variables is possible, $B(n,t)=B(t)p_{0}(n)$,
where $B(t)=B_{o}e^{\gamma t}$ is the exponentially growing bacterial
population and $p_{0}(n)$ is the normalized steady-state PDF. In
this case, Eq. \ref{eq1:a} simplifies to:

\begin{equation}
	2\gamma p_{0}(n) 
		=
	-\frac{\partial}{\partial n}(\nu(n)p_{0}(n)) 
	+ 2\gamma\intop_{n}^{\infty}\Phi(n|m)p_{0}(m)~dm.
\label{eq1:b}
\end{equation}
During cell division, daughter cells inherit proteins from mother
cells. If the partition of proteins is random, we expect that the
sharing of $m$ proteins between the two daughter cells will follow
a binary distribution. As a simplification, we treated $\Phi(n|m)$
as a $\delta$-function with $\Phi(n|m)\simeq\delta(n-m/2)$. This
approximation becomes progressively better with increasing $m$. Finally,
we follow Ref.\cite{Krishna} and assume that the protein production
rate is linear in $n$ with $\nu(n)=cn$. This means that in the absence
of cell division the proteins in an individual cell grows exponentially,
$n(t)=n_{o}e^{ct}$. With these assumptions, Eq.\ \ref{eq1:b} becomes
\begin{equation}
	2\gamma(~p_{0}(n)-2p_{0}(2n)~)=-\frac{\partial}{\partial n}(cnp_{0}(n)).
	\label{eq1:c}
\end{equation}

To appreciate the important role of cell division in redistribution
of proteins, consider a simple biologically relevant situation where
a group of cells, spanning a broad range of receptor states ($0\leq n\leq n_{max}$),
produces no protein $\nu(n)\simeq0$. This could be a result of kinetic
reasons such as deficiency in some transcriptional factor or, as discussed
below, a lack of a DNA loop necessary for the transcription initiation
of LamB\cite{Richet}. Although these cells produce no protein, they
are genetically and physiologically identical to their cousins in
the majority population and should be capable of producing a large
number of proteins if this kinetic barrier is eliminated. If this
protein production is rare and temporary, we would have a simple fire-and-divide
scenario in which a large number of LamB receptors can be generated
in a short burst. Cell division then dilutes the protein number, filling
the low $n$ states. We can account for this mechanism by adding a
term $g(n)p_{0}(0)$ to the right hand side of Eq.\ \ref{eq1:c},
where $g(n)p_{0}(0)$ represents a burst of proteins from the $n\simeq0$
state. We assume that $g(n)$ is a continuous function peaked at $n_{max}$,
which defines the upper bound of the minority population. It follows
that within $0<n<n_{max}$, the steady-state solution is $p_{0}(n)\sim n^{-1}$,
which is consistent with some of our observations in the induced minority
populations displayed in Figs.~\ref{fig:1} and \ref{fig:2}.

\begin{figure}

\includegraphics[width=8cm]{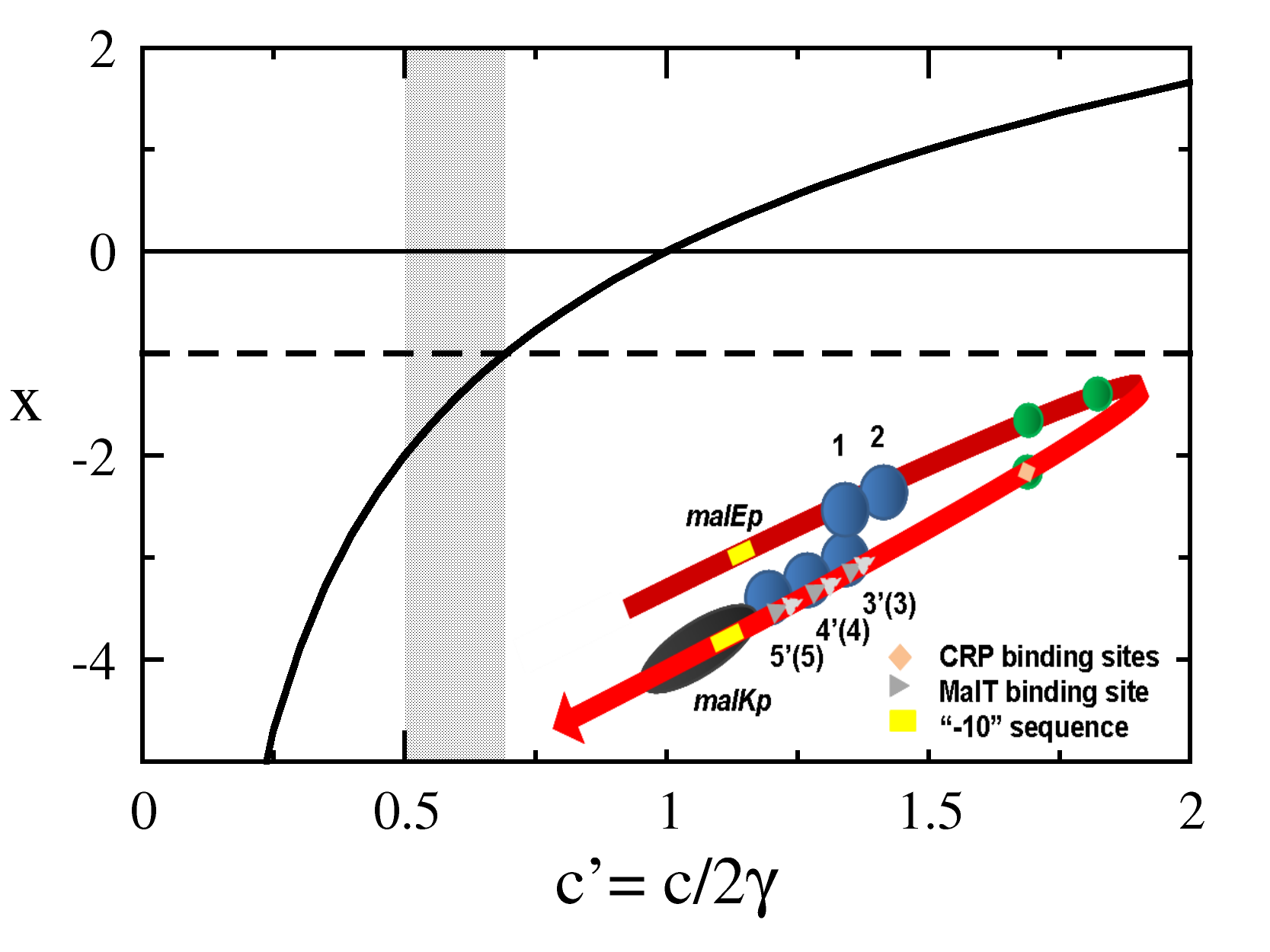}

\caption{
The solutions to the equation $2^{x+1}=1+c'(x+1)$ are given by the
dashed line ($x=-1$) and the solid curve. The observed exponent $-2<\alpha<-1$
for the minority cells grown in maltose is within the shaded area,
$0.5<c'<0.7$. The inset depicts the DNA loop for the transcription
initiation of \emph{lamB} gene. The red ribbon represents the regulatory
region of \emph{malKp}, and the big arrow is the direction of transcription
when \emph{malKp} is active. The downstream genes include \emph{lamB}
and other two proteins. The small green spheres represent CRPs that
cause the DNA to bend, the large spheres are MalTs that are bound
to sites 1-2 and 3'-5' as shown, and the spheroid represents the RNA
polymerase. Without the formation of the loop, the three MalTs adjacent
to \emph{malKp} will bind to alternative sites 3-5 that are 3 bp upstream
from the 3'-5' sites\cite{Richet}. In this case, there is no contact
between MalT and RNA polymerase, and \emph{malKp} is inactive.
\label{fig:3}
}

\end{figure}

We show below that a finite protein production rate ($c\not=0$) can
significantly alter the protein PDF, but the $n^{-1}$ distribution
remains as a solution of Eq. \ref{eq1:c}. Note that if $p_{0}(n)$
is an homogeneous function of $x$ degree, $p_{0}(n)\sim n^{x}$,
then so is $\frac{\partial}{\partial n}(cnp_{0}(n))$. Substituting
this power-law form into Eq.\ \ref{eq1:c} gives the transcendental
equation $2^{x+1}=1+c'(x+1)$, where $c'=c/2\gamma$ is the reduced
protein production rate. This equation has two roots, one being $x=-1$
independent of $c'$, and the other root being a monotonically increasing
function of $c'$, as shown in Fig.\
\ref{fig:3}. We find that the second root $x$ is less than $-1$
for small protein production rate $c'<\ln2$, indicating that $n=0$
is an accumulation point. For $c'>\ln2$, $x>-1$ with $x$ becoming
positive for $c'>1$. In this case, the cummulative probability $\intop_{0}^{n}p_{0}(n')dn'$
is infinite for large $n$. We interpret this to mean that the cells
leave the minority state for $c'>\ln2$. Fig.\ \ref{fig:3} also
shows that $x$ is a steep function of $c'$ for small $c'$. This
may be why the exponent $\alpha$ varies from run to run. For instance,
for the maltose-grown cells, the variations seen in $\alpha$ ( $-2<x<-1$)
only corresponds to a relatively small change in $c'$, i.e., $0.5<c'<0.7$. 

It is unclear at present what mechanism gives rise to the segregation
of cells into the majority and the minority populations. One possibility
is due to the DNA loop required to initiate the transcription of \emph{lamB}
(see Fig.~\ref{fig:3}), which is under control of the\emph{ malKp
}promoter\cite{Boos}. Previous genetic studies have shown that the
binding of an RNA polymerase to \emph{malKp} requires the repositioning
of three MalTs, the primary transcriptional activators, from a set
of non-productive sites (3-5) to a set of productive sites (3'-5')
that is staggered by 3 bp\cite{Richet}. The repositioning requires
the formation of a loop involving two additional MalTs at sites $\sim200$
bp apart from the first three, and the binding of CRPs (cAMP receptor
protein) to three sites located in the intervening region of the DNA
as delineated in Fig. \ref{fig:3}. Unlike \emph{lac} and \emph{ara}
operons, where the DNA loop enhances repression, here it is essential
for the initiation of \emph{lamB} transcription. If the bacteria in
the minority population cannot form the loop due to a deficiency either
in CRP or MalT proteins, LamB will be transcribed at a very low rate\cite{Richet}.
Since the formation of a loop is a statistical event subject to thermal
fluctuations, the ratio of the minority to the majority populations
may reflect the energy difference between the open and the closed
conformations. For a large population, the statistical nature of the
process ensures that the ratio between the two populations will be
constant as seen in our experiment\cite{Vilar}. Early experiments
also showed that MalT expression is limited both in the transcription
and in the translation\cite{Chapon}, indicating that maltose regulon
has evolved in a way that enhances the expression noise in the population.
This is in contrast with most regulatory proteins for which the system
is typically constructed in a way to minimize the protein-number fluctuations.
LamB, however, is exploited not only by $\lambda$ but also by several
other bacteriophage such as K10 and TP1. The noisy expression pattern
makes biological sense because the phenotype can help the bacterial
population cope with environmental fluctuations. In the uninduced
state, the repressed cells still have a broad distribution of receptors,
making them ready for full induction in the case that maltose or maltodextrins
become available. In the fully induced case, the majority of cells
can prosper in the environment but a small minority is protected from
potential phage infections. This scenario is consistent with our recent
observations that the heterogeneous bacterial population is more fit
in an environment where $\lambda$ phage is present\cite{Chapman-McQuiston}.

In conclusion, we found that when \emph{lamB} gene is fully induced
in a bacterial culture, the cells segregate into two populations with
very different LamB receptors numbers per bacterium. The majority
population, which consists of more than 98\% of bacteria, has a mean
receptor number $\sim500$ but the minority population has only a
few receptors on average. We believe that this novel phenotype is
selected for because the bacterium coevolves with viruses that exploit
its receptor. We propose that the controlling mechanism is the DNA
loop structure for transcriptional initiation of \emph{lamB} gene.
A noticeable feature of the minority population is that the number
of proteins obeys a power-law distribution $p(n)\propto n^{-\alpha}$
with $\alpha$ varying from run to run in induced cell cultures. Using
a simple model, we showed that the power-law scaling for $p_{0}(n)$
can be accounted for by a low protein production rate and a back-cascade
process due to cell division. Since the model relies on very general
features of protein production and cell division, we believe that
our result is applicable to other proteins in the minority population.

We would like to thank C.C. Chen and D. Jasnow for useful discussions.




\end{document}